# Toward Secure Web-to-ERP Payment Flows: A Case Study of HTTP Header Trust Failures in SAP-Based Systems

Vick Dini
*DEIB*
*Politecnico di Milano*
Milan, Italy
vickpierce.dini@polimi.it

*Abstract*—Electronic banking portals often sit in front of enterprise resource planning (ERP) systems such as SAP, mediating payment requests between users and back-end financial infrastructure. When these integrations place excessive trust in client-supplied HTTP metadata, subtle design flaws can arise that undermine payment integrity. This article presents a retrospective, anonymized case study of an SAP-based payment flow in which weaknesses in HTTP-level validation allowed the front-end application to incorrectly treat unpaid transactions as completed. Rather than provide a reproducible exploit, we abstract the scenario into a general vulnerability pattern, analyze contributing architectural decisions, and propose concrete design and verification practices for secure web-to-ERP payment processing. The discussion emphasizes formalizing payment state machines, strengthening trust boundaries, and incorporating regular security review into integration projects.

## I. Introduction

Online banking and web-based payment services have become central to how individuals and organizations conduct financial transactions[1]. From utility bills and tax payments to e-commerce purchases, users increasingly rely on browser-based interfaces backed by complex chains of systems. In many institutions, this chain includes an ERP platform such as SAP, which handles accounting, invoicing, and communication with external banking networks.

The widespread use of such architectures has raised important security questions. While transport-layer protections like HTTPS are now standard, many real-world failures occur not at the cryptographic layer but in the design of application-level workflows that span multiple components [1]. When web front ends make assumptions about the trustworthiness of client-side information, and when those assumptions bleed into how back-office systems record financial state, inconsistencies and exploitable gaps can arise[2].

This article examines one such incident, which occurred more than a decade ago in a production environment. A public-facing web portal was integrated with an SAP system responsible for managing payment status. Under certain conditions, the portal could be induced to treat a payment as successful, even though the underlying financial transaction had not been properly completed. The root cause lay not in a single software defect, but in the way HTTP-level metadata and session state were used as evidence of payment completion.

The incident was reported to the responsible parties and remediated at the time. Here, we revisit it purely as an instructive case study. We deliberately omit precise transaction details, message formats, and exploit steps, and instead extract more general lessons about web-to-ERP payment design. Our contributions are:

- A conceptual model of how client-controlled HTTP metadata can influence payment workflows in ERP-backed systems.
- An analysis of the architectural and process-level decisions that allowed a payment-state inconsistency to arise.
- A set of design and verification recommendations that practitioners can apply when building or auditing similar systems today.

The article is structured as follows: Section II provides background on SAP-based payment workflows and the role of HTTP in web integration; Section III introduces a system model for the kind of architecture under discussion; Section IV presents the anonymized case study at a conceptual level; Section V formalizes the threat model; Section VI analyzes root causes; Section VII describes mitigation strategies; Section VIII situates the case in the broader context of related work; Section IX reflects on ethical considerations; and Section X concludes.

## II. Background: SAP-Based Payment Workflows

SAP is widely deployed as a back-office platform for financial operations, including accounts receivable, billing, and payment reconciliation. In many organizations, direct access to SAP is restricted; customers and external users interact instead with a separate web application that exposes selected functionality. Payment workflows thus bridge at least two distinct domains:

---

[1] https://cheatsheetseries.owasp.org/cheatsheets/Third_Party_Payment_Gateway_Integration.html

[2] https://owasp.org/Top10/2021/A01_2021-Broken_Access_Control/

- The public web domain, where users authenticate, initiate actions, and view results through a browser.
- The internal ERP domain, where business objects (invoices, orders, contracts) are stored and updated, and where back-end integrations to banks or card processors reside.

A common high-level payment flow in such an environment is:

1) A user authenticates to a web portal and selects an item to pay.
2) The portal constructs a payment request and forwards it toward a payment gateway or directly to an SAP component that coordinates with the bank.
3) The bank or payment processor authorizes or declines the transaction.
4) The result is propagated back to SAP, which updates the corresponding business object and possibly sends a status indication back to the portal.
5) The portal reflects the final status to the user and, if applicable, unlocks services or content.

We illustrated a slightly more complex version of this flow in Fig. 1.

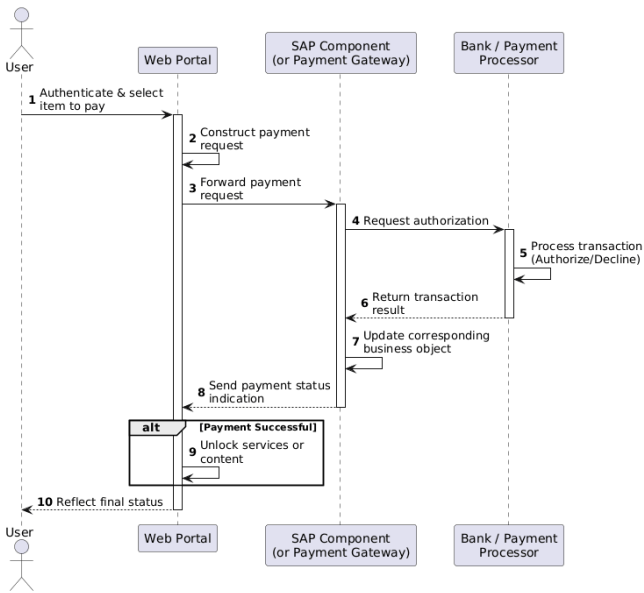

Fig. 1. SAP Payment Flow

This flow can be implemented in various ways: server-side requests from the portal to SAP, browser-redirects through third-party payment pages, asynchronous callbacks, or message queues. Regardless of the mechanism, correct behavior relies on a shared understanding of payment state and on clear trust boundaries [2] between components.

On the web side, HTTP serves as the transport protocol between browser and portal. HTTP requests and responses contain a start line, a set of headers with metadata, and an optional body[3] Headers expose information such as cookies, session identifiers, content types, and application-specific fields. Since most of these values ultimately stem from the client, they must be handled as untrusted input whenever they have security relevance.

### III. System Model

To frame the discussion, we consider a simplified model comprising four main components:

- Client browser (**C**): Operated by the user, responsible for initiating requests over HTTPS to the web portal. The client can be fully controlled by an attacker in our threat model.
- Web portal (**P**): A web application that authenticates users, displays pending invoices or orders, and orchestrates the payment process from the user's perspective.
- Payment gateway (**G**): A component (internal or external) that interacts with banking or card-network infrastructure to authorize and capture funds.
- ERP system (**E**): An SAP system that stores business objects and payment records, and communicates with **G** as necessary.

Conceptually, the payment state for a given business object **B** can be modeled as a finite state machine with states such as: *Created, Payment initiated, Authorization pending, Authorized, Captured, Settled, Failed* or *Canceled*. In a robust design, **E** maintains the authoritative state of **B**, and **P** acts as a view and control layer that issues requests and queries but doesn't independently decide the final state. The transition from *Payment initiated* to any state that confers durable benefits to the user (e.g., *Settled*, *Access granted*) should depend solely on information that flows from trusted actors (**G**, **E**), not on mutable client signals.

Implementations vary in how they encode intermediate state and how they propagate status between components. Some use server-side sessions and internal APIs. Others encode state identifiers or flags in URLs, hidden form fields, or custom HTTP headers. It is in these design choices that opportunities for inconsistency and abuse can emerge.

### IV. Case Study: A Payment-State Consistency Failure

The incident that motivates this article took place in an environment structurally similar to the model above. Users accessed a web portal over HTTPS to view pending charges and initiate payments. The portal communicated with an SAP system that in turn interfaced with the banking infrastructure. Under ordinary operation, the end-user experience was straightforward: clicking through the payment flow led to a confirmation message, and the corresponding entry in the user's account was marked as paid.

Subsequent investigation, however, revealed that the logic connecting front-end behavior to back-end payment status was more fragile than it appeared. In particular, the portal relied on a combination of:

- Session-level indicators that were tracked via HTTP mechanisms (such as cookies and other headers).
- Application-level flags that summarized the apparent outcome of the payment process.

---
[3]https://www.rfc-editor.org/rfc/rfc7230.html

- Assumptions about the ordering and structure of requests in a "normal" payment flow.

In essence, certain HTTP-level attributes, as observed by **P** in the course of a session, influenced whether it presented a payment as successful and whether it invoked back-end updates to mark the associated business object as paid. While SAP still maintained its own records, the portal's view of reality could diverge, at least for some period, from the ERP's authoritative state.

From the standpoint of an attacker controlling **C**, this arrangement created a subtle opportunity. By manipulating or replaying specific HTTP messages within an authenticated session, it was possible to trigger execution paths in **P** that corresponded to post-payment logic, without having completed a valid payment via **G** and **E**. In effect, the system's internal notion of "this user has finished paying for **B**" could be misaligned with what had actually occurred in the banking network.

The precise messages and modifications that made this possible are intentionally not described here. The focus instead is on the underlying architectural pattern: an overreliance on client-visible, mutable HTTP metadata to drive critical transitions in the payment state machine.

## V. Threat Model

To reason about the generality of this vulnerability pattern, we articulate an abstract threat model:
- The attacker controls a client device interacting with **P** and can inspect, modify, and replay HTTP requests and responses within a session.
- The attacker can use common web debugging or interception tools to alter headers, cookies, and parameters, and to re-issue requests out of their original sequence.
- The attacker has valid credentials for at least one account on the portal (e.g., their own). There is no assumption of elevated privileges within **E** or **G**.
- The attacker cannot break modern cryptographic primitives. Transport-layer security (HTTPS) functions as intended; there is no eavesdropping or modification in transit beyond the client end.
- The attacker's goal is to induce the combined system (**P** + **E**) to treat a payment as completed—or to provide the practical effects of completion—without a corresponding, legitimate financial transaction.

Within this model, attacks do not depend on low-level protocol vulnerabilities, such as TLS weaknesses. Instead, they take advantage of higher-level assumptions about how state evolves and how different components attribute meaning to HTTP-level artifacts.

## VI. Root Cause Analysis

### A. Overtrust in Client-Side Indicators

A central issue was that **P** treated certain client-visible signals as strong evidence of payment completion. For example, particular combinations of session flags and request attributes were interpreted as indicating that the user had successfully navigated the payment process. While such signals may accurately describe a typical user's path, they are not, by themselves, proof that a payment has been authorized and settled by **G** and recorded by **E**.

In secure design, these indicators should be viewed as hints or triggers to check back-end state, not as authoritative facts. The conflation of "the user appears to have returned from the payment flow in a certain way" with "the payment has definitively succeeded" is at the heart of the problem.

### B. Informal or Implicit State Machine

Another factor was the lack of an explicit, enforced state machine governing payment progression. Instead of modeling and implementing payments as transitions between well-defined states, the system relied on dispersed flags and conditional logic spread across layers. When payment logic is implicit in code paths rather than explicit in a shared model, it is easier for inconsistencies to emerge between what **P** believes and what **E** records.

An explicit state machine would have made it clearer which transitions are permissible and what evidence is required for each. For example, moving from *Payment initiated* to *Settled* might require a specific, authenticated message from **G** to **E**, and then a confirmation from **E** to **P**. Client-side behavior alone would never suffice.

### C. Encoding State in Client-Visible Artifacts

The design also reused or encoded aspects of transaction state in artifacts visible to the client, such as session-related values and other HTTP-level fields. When those artifacts later fed back into server-side decisions, the door was opened for a client-side attacker to influence state transitions. Even if the fields were not meant to be user-modifiable, their presence on the client makes them, in practice, modifiable.

This pattern isn't unique to the particular system described here. Many web applications serialize or cache state in URLs, cookies, or custom headers for convenience, without fully considering how those values might be tampered with. When such values become security-critical, the risk grows.

### D. Limited Cross-Checking with the ERP System

Finally, there was insufficient reconciliation between **P**'s view of payment completion and **E**'s authoritative records. While SAP did capture transaction outcomes, **P** could, under certain conditions, act as though payment had succeeded before or without verifying **E**'s state. Stronger coupling—such as mandatory checks against **E** before unlocking services—would have reduced or eliminated the exploitable window.

## VII. Mitigation Strategies and Design Recommendations

### A. Make the ERP or Payment Backend the Sole Authority

Payment completion should be determined exclusively from trusted, server-maintained data in **E** (and, where applicable, **G**). **P** should never treat client-side signals, including HTTP headers, cookies, or URL parameters, as sufficient evidence that a payment is complete. Instead, **P** should query **E** for the current state of the relevant business object or wait for a backend-initiated notification that **E** has moved the object into a "paid" state.

In practical terms, this often means:
- Using asynchronous callbacks or message queues from **G** to **E**.
- Implementing APIs through which **P** can ask **E**, "What is the payment status of object **B**?" rather than inferring it from the client's navigation.

### B. Define and Enforce an Explicit Payment State Machine

Implementers should model payment flows as state machines with well-defined transitions, preconditions, and postconditions. The model should be shared across components and reflected in code, configuration, and documentation.

Key elements include:
- Enumerated states (e.g., pending, authorized, captured, settled, failed).
- Clear rules for which messages cause transitions and in what direction.
- Documentation of which component is responsible for issuing each state change.
- Automated checks that reject or log unexpected transitions.

By enforcing this state machine at the ERP level, the system can ensure that no single HTTP interaction from **C** to **P** can skip intermediate steps.

### C. Treat All Client-Side Data as Untrusted

All values that traverse the client—headers, cookies, parameters, and bodies—should be validated and treated as untrusted. When these values represent identifiers for server-side state, the server should enforce that:
- The identifiers correspond to objects owned by the authenticated user.
- The current state of those objects is compatible with the requested operation.
- The request does not itself confer state transitions that should only follow from back-end events.

Where client-visible tokens must encode state, they should be cryptographically protected (e.g., signed) and designed so that tampering is detectable and rejected. Even then, signatures should usually attest to identity or authorization, not to payment completion itself.

### D. Strengthen Separation of Concerns

The user interface should primarily display and request information; it should not independently maintain or infer the truth about payment status. All irreversible decisions—such as granting access to paid services—should be conditioned on up-to-date confirmation from **E**.

Architecturally, this can be achieved by:
- Concentrating payment-finalization logic in **E** or a tightly controlled service adjacent to it.
- Having **P** act as a thin layer that requests actions and displays their results.
- Avoiding duplication of business logic across layers.

### E. Implement Reconciliation and Monitoring

Organizations should implement regular reconciliation between front-end records and ERP data. For example, periodic jobs can compare what **P** believes has been paid with what **E** records as settled. Any discrepancies should be flagged and investigated.

In addition, monitoring should capture unusual request sequences or patterns around payment endpoints, such as repeated partial flows, unexpected parameter values, or inconsistent transitions. While such monitoring may not prevent all abuse, it can greatly reduce dwell time and impact.

### F. Conduct Security Reviews of Integration Points

Integration projects between web front ends and ERP systems are natural points of complexity. Security reviews should include:
- Threat modeling sessions that explicitly consider client-side manipulation.
- Code and configuration reviews focusing on how payment state is represented and updated.
- Testing that attempts to replay or alter HTTP interactions at various points in the flow.

These reviews are particularly important when third-party vendors [3] or integrators implement parts of the system, as assumptions can be lost or miscommunicated across organizational boundaries.

## VIII. Related Work

The issues highlighted in this case study intersect with several strands of prior work in web security and financial systems. Research on web application vulnerabilities has long emphasized the dangers of trusting client-side parameters and session artifacts, showing how attackers can manipulate them to access unauthorized resources or bypass checks [4], [5]. Work on payment systems has documented failures arising from mismatches between web-layer assumptions and back-end settlement processes.

Similarly, the security of ERP systems, including SAP, has been the subject of numerous technical advisories and academic analyses, particularly with respect to misconfigurations, insecure interfaces, and weak trust boundaries between modules [6], [7], [8]. Case studies of high-profile payment breaches have also illustrated how complex integration chains can obscure responsibility for state management and validation [9], [10], [11].

Rather than reproduce specific findings, this article aligns with that broader body of work by focusing on one representative pattern: over-reliance on HTTP-level metadata and implicit state in a heterogeneous payment architecture. The lessons drawn here reinforce calls in the literature for explicit state machines, clear authority boundaries, and rigorous validation of all data crossing trust boundaries[4] [12].

---

[4]https://owasp.org/www-project-web-security-testing-guide/

## IX. Ethical Considerations

Publishing analyses of real security incidents raises important ethical questions. In this case, several measures have been taken to minimize risk and respect stakeholders:

- The incident is historic and was reported to the relevant parties at the time. The description here is abstracted to avoid identifying organizations, individuals, or specific systems.
- Concrete exploit steps, exact HTTP messages, and configuration details are deliberately omitted. The goal is to illuminate a general design flaw, not to enable replication of past or current attacks.
- The discussion focuses primarily on architectural lessons and mitigations that can help others avoid similar issues.

## X. Conclusion

The case study presented in this paper illustrates how subtle, application-level decisions at the boundary between web portals and ERP systems can lead to significant security vulnerabilities in payment processing. By treating certain HTTP-level indicators as evidence of payment completion, and by lacking an explicit, enforced payment state machine, the system in question exposed a gap between perceived and actual financial state.

The broader lesson is that secure design in complex financial architectures demands clear trust boundaries, explicit state modeling, and a disciplined approach to the use of client-side data. Even when an ERP platform such as SAP is regarded as a trusted system of record, its extensive interfaces and integrations with external applications can expose additional attack surfaces and vulnerabilities if those connections are not rigorously secured. ERP systems or dedicated payment backends should act as the sole authority on payment completion, while web front ends should defer to their records rather than infer final state from browser interactions.

As organizations modernize legacy systems and build new integrations, the patterns and recommendations described here can serve as a checklist for avoiding similar pitfalls. Future work may include formal verification of payment state machines, tooling to automatically detect misuse of client-originated metadata in critical workflows, and more systematic documentation of integration-layer vulnerabilities across different ERP platforms.